\documentclass[aps,prl,preprint,groupedaddress,showpacs]{revtex4}
\usepackage{graphicx}% Include figure files
\usepackage{dcolumn}% Align table columns on decimal point
\usepackage{bm}% bold math
\begin{document}
\title{Observation of coherence revival and fidelity saturation in a $\delta$-kicked rotor potential}
\author{Saijun Wu}
%\altaffiliation[Current electronic address: ]{saijunwu@nist.gov}
\altaffiliation[Current address: ]{NIST, Gaithersburg, MD 20899}
\author{Alexey Tonyushkin}
\author{Mara G. Prentiss}
%\altaffiliation[Electronic address: ]{prentiss@fas.harvard.edu}
\affiliation{Department of Physics, Harvard University, Cambridge,
MA, 02138}

\date{\today}

\begin{abstract}
We experimentally investigate the effect of atomic $\delta$-kicked
rotor potentials on the mutual coherence between wavepackets in an
atom interferometer. The differential action of the kicked rotor
degrades the mutual coherence, leading to a reduction of the
interferometry fringe visibility; however, when the repetition rate
of the kicked rotor is at or near the quantum resonance, we observe
revival of matter-wave coherence as the number of kicks increases,
resulting in non-vanishing coherence in the large kick number limit.
This coherence saturation effect reflects a saturation of fidelity
decay due to momentum displacements in deep quantum regime. The
saturation effect is accompanied with an invariant distribution of
matter-wave coherence under the kicked rotor perturbations.
\end{abstract}

\pacs{05.45.Mt, 32.80.Lg, 39.20.+q, 03.75.dg}

\maketitle
\section{}

Information about the stability of quantum evolution in response to
small perturbations is of both fundamental and practical importance.
At a fundamental level, the study of quantum stability elucidates
how quantum irreversibility emerges from the unitary wave evolution
governed by the Schr{\"o}dinger equation~\cite{StabilityPeres}. At a
technological level, quantum control is implementable only if the
quantum trajectories are stable against small parameter errors.
Quantum stability is best characterized by a decay of fidelity which
gives the squared overlap between a perturbed quantum trajectory
with its unperturbed copy~\cite{StabilityPeres}. In contrast with
the generic Lyaponov decay of fidelity characteristic of classically
chaotic systems~\cite{chirikov79, classicalecho04}, corresponding
quantum systems can survive certain types of perturbations over long
time, leading to quantum freeze or saturation of fidelity
decay~\cite{ FidelityDecayReview06, SaturationFidelity06,
phasespaceDispDiego}. Quantum fidelity decay is practically related
to interferometry experiments~\cite{NMRecho95, BilliardsDavidson,
Billiardsecho,SignatureStability03}, where the perturbation under
investigation differentially perturbs two quantum trajectories in
the interferometer and reduces the overlap at the interferometry
output. We shall generally consider a beamsplitter operation $\hat
S\sim 1+\hat D$ applied at $t=0$ that puts an atom into a
superposition of two quantum trajectories at $0<t<T$: the
``original'' trajectory $e^{-i \hat H_0 t}|\psi\rangle$ and the
``shifted'' trajectory $e^{-i \hat H_0 t}\hat D |\psi\rangle$ where
$\hat H_0$ is a perturbation-free Hamiltonian. With additional
interferometry operations (mirrors and beamsplitters) to create an
effective time-reversal, the two trajectories can be optimally
overlapped at the interferometry output. The perturbation $\hat V$
applied at $0<t<T$ can be written in the frame of the ``original''
and ``shifted'' states as $\hat V_1=\hat V$ and $\hat V_2=\hat
D^{-1} \hat V \hat D$ respectively. The trajectory overlap $A$ at
the interferometry output is expected to reduce to $\tilde A$ due to
the differential perturbation. The ratio $\tilde A/A$ gives the
fidelity amplitude $f=\langle \psi| \hat U_2^{-1}(T) \hat
U_1(T)|\psi\rangle$ with $\hat U_{1,2}(T)= \hat{\mathrm T}e^{-i
\int_0^T \hat V_{1,2} {\rm d} t}$ in the interaction picture
($\hat{\mathrm T}$ the time-ordering
operator)~\cite{FidelityDecayReview06}.

Much of the earlier experimental work studied the fidelity decay
with internal-state echoes~\cite{NMRecho95, BilliardsDavidson,
Billiardsecho,SignatureStability03} where $\hat D$ is associated
with an internal-state operation and $\hat V_{1,2}$ correspond to
perturbations in different internal states. In particular, a Ramsey
interferometer was suggested in ref.~\cite{MotionStabilityRaizen} to
study the stability of an atomic $\delta$-kicked rotor (ADKR), which
is an atom-optics realization of a $\delta$-kicked rotor by
subjecting cold atoms to periodic ``kicks'' from an optical standing
wave (SW) pulse train~\cite{kickedrotor95}. Classical dynamics of a
$\delta$-kicked rotor is described by the standard map, which
displays generic features of classical
instabilities~\cite{chirikov79}. Quantum dynamics of ADKR is
strongly affected by two relevant frequencies, the repetition rate
of the pulses and the atomic recoil frequency. A quantum resonance
happens when the two frequencies coincide, where dynamics of atoms
becomes deeply quantum mechanical due to matter-wave interference
and integrability~\cite{kickedrotor95, SaturationFidelity06,
highorder06}.

This work studies fidelity decay of matterwaves in an ADKR potential
under a momentum displacement~\cite{phasespaceDispDiego}. The
displacement operation is experimentally generated by a grating
diffraction due to the atomic recoil effect in an atom
interferometer~\cite{AIBerman, PrichardInf, TLCahn97}. We observe
loss and revival of matter-wave coherence under a few ADKR kicks
which saturates to a constant value insensitive to hundreds of kicks
at quantum resonance. Our observation for the first time
demonstrates stability of matterwaves in a classically-chaotic ADKR
potential with a saturation of fidelity
decay~\cite{SaturationFidelity06}. By introducing a ``displacement
diagram'' to organize the total mutual coherence for wavepacket
pairs in the interferometer, we provide an external-state extension
of the formula developed in
ref.~\cite{MotionStabilityRaizen,SaturationFidelity06} and show
excellent agreements between our observation and the theory. We also
show that the saturation of fidelity is accompanied with an
invariant distribution of matter-wave coherence which suggests
interferometric applications of ADKR.

\begin{figure}
\centering
\includegraphics [width=3.5 in,angle=0] {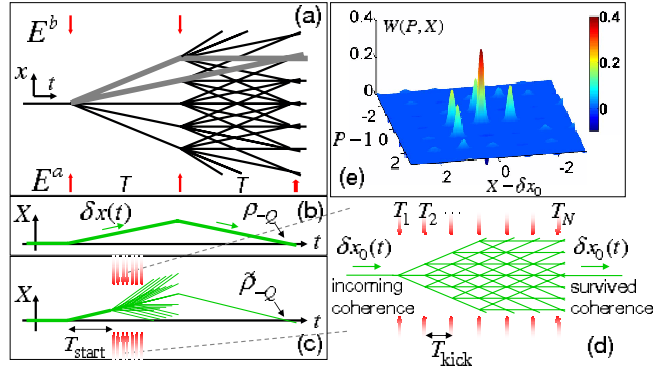}
\caption{(Color online) (a): Recoil diagram for a 3-pulse
interferometer. (b): Diagram for the relative displacement between
two interfering wavepackets of interest. (c): Displacement diagram
of (b) perturbed by ADKR. (d): An expanded view of the interaction
regime in (c) in a moving frame. Here $\delta x_0(t)$ corresponds to
$\delta x (t)$ in (b) with $0<t<T$. (e): 2D plot of a typical {\it
invariant} distribution of $W(P,X)$ under ADKR around $P=\hbar
(Q'=Q)$ and $X =\delta x_0(T_i)$ at quantum resonance ($P$ and $X$
is normalized to $\hbar Q$ and $\frac{\hbar Q}{m}T_{\rm kick}$
respectively). The incoming coherence $W(P,X)$ is a Gaussian
distribution with unity amplitude. The central peak corresponds to
the survived coherence, given by Eq.~(\ref{fidelityNi}) with
$\phi=1.6$.}\label{fig1}
\end{figure}

Our interferometry scheme follows those developed in
ref.~\cite{TLCahn97}. An optical SW is pulsed to create a sinusoidal
light shift potential, that diffracts atoms into multiple
diffraction orders, with $n^{th}$ diffraction order weighted by
amplitude $i^n J_n(\Theta)$. Here $\Theta$ is the interferometer SW
pulse area, and $J_n$ the $n^{th}$ order Bessel function. Successive
applications of two pulses at $t=0$ and $t=T$ leads to a revival of
atomic density grating at around time $t =2 T$. The revived atomic
density grating has a k-vector ${\bf Q}={\bf k}^a-{\bf k}^b$, that
is the same as the k-vector of the interferometer SW which is
composed of traveling light fields $E^{a,b}$ with k-vectors ${\bf
k}^{a,b}$ [red/gray arrows in Fig.~\ref{fig1}(a)]. A ``grating
echo'' technique~\cite{gratingechoOrigin, TLCahn97} that monitors
the Bragg scattering of light from the $E^a$ mode into the $E^b$
mode can retrieve the Fourier component of the atomic density
grating $\rho_{-Q}$.

The matter-wave diffraction paths in the interferometer are sketched
with a $x-t$ recoil diagram~\cite{billiardball93} in
Fig.~\ref{fig1}(a), where straight lines guide the centers of the
diffracted wavepackets. The atomic density fringes at around time
$t=2T$ are due to the interference between pairs of wavepackets
along adjacent diffraction paths that form ``triangle loops''.
Obviously, the relative displacements between the two diffraction
paths in all these triangles are the same, due to the quantized
recoil momentum $\hbar Q$ from the interferometer SW field. For a
convenient discussion of the total mutual coherence between all
these wavepacket pairs, we consider a ``displacement diagram'', that
only marks the relative displacement $\delta  x(t)$ between these
wavepacket pairs in a $X-t$ coordinate. The total mutual coherence
between the wavepacket pairs induces a peak in the phase-space
correlation function $W(P, X, t)={\rm Tr}[\hat \rho(t) \hat D(P,
X)]$ at $P=\pm \hbar Q$ and $X=\delta x(t)=\int_0^t \frac{P}{m} {\rm
d}t $. Here $\hat \rho(t)$ is the single atom density matrix
operator, $\hat D( P, X)=e^{i/\hbar ( P \cdot \hat x - X \cdot \hat
p)}$ is a displacement operator ($\hat x$ and $\hat p$ is the
position and momentum operator respectively)~\cite{foot:exp1}. The
$\delta x(t)$ in Fig.~\ref{fig1}(b) will be referred to as
``displacement lines'' that guide peaks of the correlation function
$W(P,X,t)$ in the displacement diagram. Using the properties of
$\hat D(P,X)$, it is easy to show that the interaction of an impulse
of SW $ V(\hat x, t)=\hbar \theta \cos (Q' \hat x) \delta (t)$ (with
k-vector $Q'$) leads to
\begin{equation}
W(P,X,0^+)=\sum_{n}J_n \left(2\theta \sin\frac{Q' X}{2}\right)W(P-n
\hbar Q',X,0^-)\label{weylrule2}.
\end{equation}
Thus the peaks of $W(P,X,t)$ can be shifted by multiples of $\hbar
Q'$ along the $P$ axis with a SW ``diffraction''. Also, notice that
an interference fringe is expected whenever the displacement line
intersects the $X=0$ axis, since the associated wavepacket pairs
overlap in real space. In particular, we have the interferometer
output $\rho_{-Q}(2 T)=W(P=-\hbar Q,X=0,t=2 T)$
[Fig.~\ref{fig1}(b)].

We now consider the same interferometry sequence represented by
Fig.~\ref{fig1}(a),~\ref{fig1}(b) to be perturbed by an ADKR
potential, that is composed of $N$ pulses of SW at a repetition rate
of $1/T_{\rm kick}$, starting at $t=T_{\rm start}$ [red/gray arrows
in Fig.~\ref{fig1}(c), (d)]. We consider the kicked rotor SW with a
grating k-vector $Q'$ close but not exactly equal to $Q$, that is
realized by introducing a small intersection angle between the
kicked rotor SW and interferometer SW. The ADKR Hamiltonian is given
by $H(\hat p, \hat x, t)=\hat p^2/2m +V_{\rm KR}(\hat x, t)$ with
\begin{equation}
V_{\rm KR}(\hat x, t)=\hbar \theta \cos (Q^{\prime} \hat x) \sum
\nolimits _{i=1}^{N} \delta(t-T_i).
 \label{rotorpotential}
\end{equation}
Here $T_i=T_{\rm start}+(i-1) T_{\rm kick}$ specifies $\{T_i\}$. The
impact of ADKR generally leads to a reduction of interferometry
fringe contrast from $\rho_{-Q}(2 T)$ to $\tilde \rho_{-Q}(2 T)$,
due to the differential perturbation to the pairs of wavepackets. We
define a dephasing factor $f=\tilde \rho_{-Q}(2 T)/ \rho_{-Q}(2 T)$
to characterize the loss of the total mutual coherence. To express
$f=f(\theta,\{T_i\})$ analytically we consider the displacement
diagram in Fig.~\ref{fig1}(c), with the interaction part expanded in
Fig.~\ref{fig1}(d) in a frame co-moving with $\delta x_0(t)$. Here
$\delta x_0(t)=v_Q t$ corresponds to the unperturbed displacement
line at $0<t<T$ in Fig.~\ref{fig1}(b). The matter-wave coherence
contributing to the interferometry fringe is specified by $W(P,X)$
along $X=\delta x_0(t)$. Thus $f(\theta,\{T_i\})$ gives the fraction
of the coherence that survives the ADKR perturbation in a network of
displacement lines [Fig.~\ref{fig1}(c), (d)], where peaks of
$W(P,X)$ are scattered [due to Eq.~(\ref{weylrule2})] and interfere
at the vertex $(P^{(r)},X^{(s)})$ in the moving frame. Here $r$, $s$
are two integers, $P^{(r)}=r\hbar Q^{\prime}$ and $X^{(s)}=
s\frac{\hbar Q^{\prime}}{m} T_{\rm kick}$. By iteratively applying
Eq.~(\ref{weylrule2}) at each vertex, we have the dephasing factor
$f$,
\begin{equation}
f(\theta, \{T_i\})  = {\sum\limits_{\{n_i\}}}^{\prime}
{\prod\limits_{i = 1}^N {J_{n_i } \left(2\theta \sin
\frac{Q'}{2}X_i^{\{ n_i \} } \right)} }.\label{fidelity}
\end{equation}
Here $X_i^{\{n_i\}}-\delta x_0(T_i) \in \{X^{(s)}\}$, and $n_i$ is
the displacement diffraction order [Eq.~(\ref{weylrule2})] at time
$T_i$. A set of $\{n_i\}$ specifies a particular diffraction path in
Fig.~\ref{fig1}(d) while the sum in Eq.~(\ref{fidelity}) is for all
the paths satisfying $\sum n_i =r_N=0$, $\sum (i - 1)n_i = s_N=0$.

We found it is particularly simple to analytically evaluate
Eq.~(\ref{fidelity}) at quantum resonance~\cite{kickedrotor95}, with
$T_{\rm kick}=n\pi/\omega_{Q'}$ so that $X_i^{(n_i)}=\delta
x_0(T_i)+2 s n\pi /Q^{\prime}$, where $n$ is an integer,
$\omega_{Q^{\prime}}=\hbar {Q^{\prime}}^2/2 m$ is the two-photon
atomic recoil frequency and $m$ is the atomic mass. By ignoring the
negligible difference between $\omega_{Q'}$ and $\omega_Q$, we found
\begin{equation}
f_{\rm QR} (\phi ,N) = {1 \over {2\pi }}\int _{ - \pi }^\pi  {J_0
\left(\phi {{\sin (Ny/2)} \over {\sin (y/2)}}\right)} dy,
\label{fidelityNi}
\end{equation}
where $f_{\rm QR}(\phi, N)$ gives $f=f(\theta,\{T_i\})$ when $T_{\rm
kick}$ meets the quantum resonance and $\phi=2\theta \sin(\omega_Q
T_{\rm start})$ is considered as the external-state counterpart of
the differential perturbation strength for two internal
states~\cite{MotionStabilityRaizen, SignatureStability03,
SaturationFidelity06}. Equation~(\ref{fidelityNi}) is remarkable: as
suggested in ref.~\cite{SaturationFidelity06}, in the large $N$
limit $f_{\rm QR} = \int _{ - \pi /2}^{\pi /2} {J_0 [\phi
\sec(\alpha) /2 ]^2 } d\alpha/\pi>0$. This means that a significant
fraction of matter-wave coherence survives even after an infinite
number of kicks, which requires a freeze or saturation of fidelity
decay under the differential perturbation for majority of
wavepackets in the interferometer~\cite{phasespaceDispDiego,
SaturationFidelity06}.

The experimental setup is similar to that in
ref.~\cite{movingguide}. Approximately $10^7$ laser-cooled $^{87}$Rb
atoms in their ground state $F$ = 1 hyperfine level are loaded into
a magnetic guide oriented along $\bf e_x$, resulting in a
cylindrically-shaped atom sample 1~cm-long and 170~$\mu$m-wide at
25~$\mu$K temperature. The interferometer SW, with k-vector
precisely aligned along $\bf e_x$, is formed by two traveling waves
$E^a$ and $E^b$, detuned 120~MHz to the blue of the $F$=1 -
$F^{\prime}$=2 D2 transition. The interferometer SW is pulsed for
300~ns at $t = 0$ and $t = T$, with typical pulse areas of $\Theta
\sim 1.5$ . The total interrogation time $2 T$ is chosen to be
6.066~ms or 12.165~ms in different experimental trials. The ADKR
pulses are delivered by a different standing wave that is formed by
retro-reflecting a traveling laser beam that is 6.8~GHz detuned to
the red side of $F$=1 - $F^{\prime}$=2 D2 transition, and is 40~mrad
misaligned from the $\bf e_x$ direction. This SW field is pulsed at
$t = T_{i}$ according to Eq.~(\ref{rotorpotential}), with 400~ns
duration and has a typical pulse area $\theta\sim 0.1-1.3$. In
ref.~\cite{movingguide} we have shown that the magnetic confinement
introduces negligible perturbation to the matter-wave interference
along $\bf e_x$. Here we take advantage of the confinement to
maintain the 170~$\mu$m transverse atomic sample distribution across
the 2~mm-diameter ADKR laser beam, which enables a consistent ADKR
interaction strength with up to $N=150$ kicks. The ``grating echo''
signal amplitude is retrieved at around $t = 2 T$ using a
heterodyned technique~\cite{gratingechoOrigin, TLCahn97}, recorded
in repeated experiments as a certain parameter in ADKR is scanned.
We normalize the grating echo amplitude with a reference signal
level when no ADKR pulse is applied. The normalized amplitude
$f=f(\theta,\{T_i\})$ corresponds to the dephasing factor in
Eq.~(\ref{fidelity}) in the ideal model. Due to the misalignment,
the projection of ADKR k-vector along the interferometer SW
direction $\bf e_x$ gives $(\omega_{Q}- \omega_{Q'})/\omega_Q \sim
1.6 \times 10^{-3}$ that will be ignored in this paper. We normalize
$T_{\rm start}$, $T_{\rm kick}$ with respect to
$2\pi/<\omega_{Q'}>=66.4$~$\mu$s for a convenient comparison between
theory and experiment.

\begin{figure}
\centering
\includegraphics [width=3.4 in,angle=0] {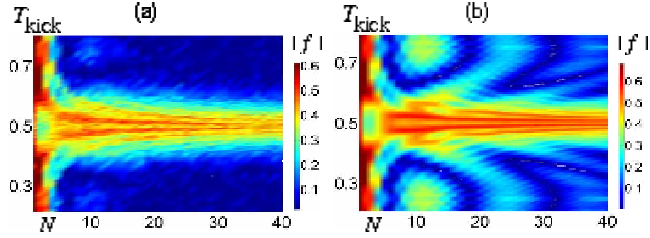}
\caption{(a): Experimentally observed dephasing factor $|f|$ with
$N$=1,...,40. Here (normalized) $T_{\rm start}=0.25+T_{\rm kick}$.
(b): Simulation of $|f|$ using Eq.~(\ref{fidelity}) with
$\theta=0.7$.}\label{fig3}
\end{figure}

The distinct behavior of interferometer contrast decay at different
$T_{\rm kick}$ with increasing $N$ in ADKR is shown in
Fig.~\ref{fig3}(a). The kicked rotor SW pulse area was chosen to be
$\theta=0.7$, calibrated with independent measurements. The density
plot in Fig.~\ref{fig3}(a) composes $40\times 80$ data points from
repeated experiments, which scan $T_{\rm kick}$ from 0.21 to 0.8,
and $N$ from 1 to 40. From Fig.~\ref{fig3}(a) we read the dephasing
factor $f$ according to the color coding, along the $N$ axis at
different $T_{\rm kick}$. For $T_{\rm kick}$ far from 0.5, a rapid
decay from $f=1$ ($N$=0) down to zero (noise level) is observed.
However, we see multiple ``bright fringes'' of high-$f$ ``flow'' to
the vicinity of $T_{\rm kick}=0.5$ with increasing $N$. At $T_{\rm
kick}=0.5$, we see a quick loss of coherence from $f=1$ to $f\sim
0.25$ at $N\sim 3$, which partly revives to $f\sim 0.45$ at $N\sim
5$, and then maintains the constant value without noticeable decay.
In Fig.~\ref{fig3}(b) we plot the results of simulation based on
Eq.~(\ref{fidelity}). We found a very good match between
Fig.~\ref{fig3}(a) and (b) both for small $N$, and for $T_{\rm
kick}$ around 0.5~\cite{foot:exp2}. The discrepancy regimes in
Fig.~\ref{fig3}(a) and \ref{fig3}(b) is likely related to increased
residual magnetic field perturbations due to atomic momentum
diffusion~\cite{kickedrotor95}, which is currently under
experimental and theoretical investigations.

\begin{figure}
\centering
\includegraphics [width=3 in,angle=0] {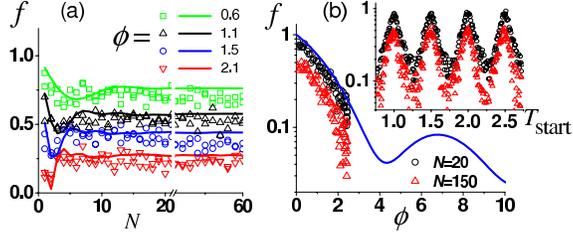}
\caption{(Color online) Dephasing factor $f$ at $T_{\rm kick}=0.5$.
Scatter plots give experimental data. Solid lines are calculated
according to Eq.~(\ref{fidelityNi}). (a): $f$ vs $N$. (b): $f$ vs
$\phi$. Inset gives the same experimental data plotted vs $T_{\rm
start}$.}\label{fig4}
\end{figure}

We accurately determine the kicked rotor SW recoil frequency
$\omega_{Q'}$ by minimizing the ADKR induced dephasing effect with
$T_{\rm kick}$ at various $T_{\rm start}$~\cite{goodtime}. Here we
fix $T_{\rm kick}=0.5$ to be at quantum resonance. According to
Eq.~(\ref{fidelityNi}), at quantum resonance a convenient parameter
to characterize the differential perturbation is $\phi=2\theta \sin
\omega_Q T_{\rm start}$. We fix $\theta$ and use $T_{\rm start}$ to
control $\phi$, and study the evolution of the dephasing factor $f$
at an increasing number of kicks. The scatter plot in
Fig.~\ref{fig4}(a) gives four measurements of this type, up to
$N=60$. For comparison, we plot the theoretical expectation with
solid lines, calculated according to Eq.~(\ref{fidelityNi}). Here in
the calculation $\theta=1.22(5)$ is determined consistent with
$\theta=0.7$ in Fig.~\ref{fig3} according to a known kicked rotor SW
intensity ratio calibrated with a photodiode. From
Fig.~\ref{fig4}(a) we see a loss, revival, and saturation of the
dephasing factor $f$ with increasing $N$ at all perturbation
strengths $\phi$. The transient feature on the loss and revival of
coherence happens more rapidly at larger $\phi$. In case $\phi=2.1$,
the dephasing factor $f$ approximately settles to the saturation
value in only $N=3$ kicks.

To study the dephasing factor $f$ vs the differential perturbation
strength $\phi$ in the saturation regime, we fix $\theta=1.22$ and
scan $T_{\rm start}$ at a fixed large number of kicks. Typical
results are plotted in the inset of Fig.~\ref{fig4}(b). We further
plot $f$ directly with $\phi$ in Fig.~\ref{fig4}(b) with up to
$\phi=2.44$, limited by ADKR pulse strength in this experiment.
Notice at $N=150$, $f$ only uniformly shifts downward by $\sim 40\%$
as compared with $f$ at $N=20$, indicating an independent coherence
loss mechanism~\cite{goodtime}. From Fig.~\ref{fig4}(b) we see an
excellent agreement between the observed dephasing factor in the
saturation regime, and those calculated according to
Eq.~(\ref{fidelityNi}) at large $N$ limit.

We come back to Fig.~\ref{fig1}(d) and Eq.~(\ref{fidelity}). Due to
the kicked rotor pulses, matter-wave coherences specified by peaks
of $W(P,X)$ are scattered and transported among 2D sites
$(P^{(r)},X^{(s)})$. The saturation of matter-wave coherence, as
predicted by Eq.~(\ref{fidelityNi}) and confirmed in
Fig.~\ref{fig4}, requires a distribution of $W(P,X)$ among these
sites {\it invariant} under the ADKR actions. Figure~\ref{fig1}(e)
gives one example of the invariant distribution, calculated with an
extension of Eq.~(\ref{fidelity}) with general $r_N$, $s_N$ at large
$N$ limit. The invariance of the 2D matter-wave coherence
distribution under the ADKR action compliments the stable momentum
population distribution when ADKR meets quantum
resonance~\cite{kickedrotor95}, and is clearly an interference
effect in the ADKR scattering network [Fig.~\ref{fig1}(d)]. By
modifying the interferometry scheme to probe atomic fringes at
$t=2T\pm T_{\rm kick}$, we have experimentally observed the
coherence peaks displaced by $\pm v_Q T_{\rm kick}$ along the $X$
axis. The peaks displaced along the $P$ axis by multiples of $\hbar
Q$ can instead be probed by a ``stimulated echo''~\cite{TLCahn97}.
The invariant distribution of coherence may be explored for robust
manipulation of matter-waves in phase space for interferometric
applications.

In conclusion, we have observed a revival and saturation of
matter-wave coherence with increasing number of ADKR kicks at or
near the quantum resonance with an atom interferometer. Our
experiments demonstrate that the wave interference rescues the
classical instability of ADKR in the deep quantum regime where
semi-classical pictures fail~\cite{phasespaceDispDiego,
SaturationFidelity06}, in excellent agreement with an external-state
extension of fidelity decay theory discussed
recently~\cite{SaturationFidelity06}. The ADKR in this work already
corresponds to a classical Hamiltonian deep in the chaotic
regime~\cite{chirikov79, kickedrotor95}. Wave stability at or near
quantum resonances in classically chaotic ADKR stems from the
stability of the corresponding pseudo-classical
map~\cite{QAM02,SignatureStability03}. By installing a chirped
kicked rotor SW, we plan to explore stable structures in the
pseudo-classical map due to Quantum Accelerator Modes~\cite{QAM02,
SignatureStability03, SaturationFidelity06}, and search for
invariant distribution of matter-wave coherence for precision
measurements.

\begin{acknowledgments}
We thank E.~J.~Su for contributions to the interferometry setup, and
C.~Petitjean and E.~J.~Heller for very helpful discussions. This
work is supported by MURI and DARPA from DOD, ONR and U.S.
Department of the Army, Agreement Number W911NF-04-1-0032, by NSF,
and by the Charles Stark Draper Laboratory.
\end{acknowledgments}

%\bibliographystyle{apsrev}
%\bibliography{movingguide}

\begin{thebibliography}{99}
\bibitem{foot:exp2}
The observation is confirmed at $T_{\rm kick}=n/2$ with $n>1$.
\end{thebibliography}

\end{document}